\documentclass[showpacs,twocolumn,superscriptaddress]{revtex4}
\UseRawInputEncoding
\usepackage{doi}%<----------
\usepackage{hyperref}
\hypersetup{
  colorlinks=true,        % false: boxed links; true: colored links
  linkcolor=blue,         % color of internal links
  citecolor=cyan,         % color of links to bibliography
}

\usepackage[utf8x]{inputenc}
\DeclareUnicodeCharacter{2212}{\textendash}
\usepackage{graphicx}% Include figure files
\usepackage{dcolumn}% Align table columns on decimal point
\usepackage{bm}% bold math
\usepackage{color}

\usepackage{amsmath}
\usepackage{amssymb}
\usepackage{mathrsfs}

\begin{document}

\title{Constraints on the magnetized Ernst black hole spacetime through quasiperiodic oscillations 
}

\author{Sanjar Shaymatov}
\email{sanjar@astrin.uz}
\affiliation{Institute for Theoretical Physics and Cosmology, Zheijiang University of Technology, Hangzhou 310023, China}
\affiliation{Akfa University,  Milliy Bog Street 264, Tashkent 111221, Uzbekistan}
\affiliation{Ulugh Beg Astronomical Institute, Astronomicheskaya 33, Tashkent 100052, Uzbekistan} \affiliation{National University of Uzbekistan, Tashkent 100174, Uzbekistan} 
\affiliation{National Research University TIIAME, Kori Niyoziy 39, Tashkent 100000, Uzbekistan}
\affiliation{Tashkent State Technical University, Tashkent 100095, Uzbekistan}

\author{Mubasher Jamil}\email{mjamil@sns.nust.edu.pk
}
 \affiliation{School of Natural Sciences, National University of Sciences and Technology, Islamabad 44000, Pakistan}

\author{Kimet Jusufi} \email{kimet.jusufi@unite.edu.mk
}
\affiliation{Physics Department, State University of Tetovo,
Ilinden Street nn, 1200, Tetovo, North Macedonia}

\author{Kazuharu Bamba} \email{bamba@sss.fukushima-u.ac.jp
}
\affiliation{Faculty of Symbiotic Systems Science
  Fukushima University, Fukushima 960-1296, Japan}

\date{\today}
\begin{abstract}
We study the dynamics of test particles around a
magnetized Ernst black hole considering its magnetic field in the environment surrounding the black hole. We show how its magnetic field can influence on the dynamics of particles and epicyclic motions around the black hole. Based on the analysis we find that the radius of the innermost stable circular orbit (ISCO) for both neutral and charged test particles and epicyclic frequencies are strongly affected as a consequence of the impact of magnetic field. We also show that the ISCO radius of charged particles gets decreased rapidly. It turns out that the gravitational and the Lorentz forces of the magnetic field are combined, thus strongly shrinking the values of the ISCO of charged test particles. Finally, we obtain the generic form for the epicyclic frequencies and selected three microquasars with known astrophysical quasiperiodic oscilations (QPOs) data to constrain the magnetic field. We show that the magnetic field is of the order of magnitude $B\sim  10^{-7}$ Gauss, taking into account the motion of neutral particles in circular orbit about the black hole.

\end{abstract}
\pacs{%04.70.Bw, 04.20.Dw
} \maketitle

%\section{Introduction}

\section{Introduction}
\label{introduction}

In general relativity (GR), gravitational collapse of massive star at the final state of the evolution can be regarded as the fundamental mechanism for the formation of astrophysical black holes. Therefore they have been very attractive and intriguing objects due to their remarkable gravitational, thermodynamical and astronomical properties. The electromagnetic field still remains one of the interesting aspects of astrophysical black holes. In GR, gravitational collapse of massive object can be decayed with $t^{-1}$ \cite{Ginzburg1964,Anderson70}, thus suggesting that black hole has no own magnetic field. However, external factors come into play for the presence of magnetic field, i.e. the accretion discs surrounding rotating black holes~\cite{Wald74} or neutron stars~\cite{Ginzburg1964,Rezzolla01,deFelice03}. With this in view, it is believed that black holes are surrounded by magnetic field in an astrophysical context. However, such a magnetic field can be referred to as test field, i.e. $B\ll B_{max}$ that does not modify spacetime geometry \cite[see, e.g.][]{Frolov10,Aliev02,Abdujabbarov10,Shaymatov14,
Jamil15,Tursunov16,Hussain15} addressing the property of test field analytically and numerically for given background spacetime.  
So far the magnetic field was measured to be of order $\sim 10^{8}$ Gauss around stellar mass black holes and $\sim 10^{4}$ Gauss around supermassive black holes, respectively (see for example \cite{Piotrovich10}). It was also estimated at the horizon (see for example) \cite{Eatough13,Shannon13}). Following that it has been found that the magnetic field can be between $200$ and $8.3 \times 10^{4}$~ Gauss at 1 Schwarzschild radius~\cite{Baczko16} and about $B\sim 33.1 \pm 0.9~\rm{Gauss}$ in the
corona as a consequence of observational analysis of binary black hole system $V404$ Cygni ~\cite{Dallilar2018}. Recently it has been estimated to be of order $B\sim (1-30)$ {Gauss} as a result of analysis of imaged polarized emission around the supermassive black hole in M87 under collaboration of Event Horizon Telescope (EHT)~\cite{EventHorizonTelescope:2021srq,EventHorizonTelescope:2021bee}.  
However, those estimated values still remain candidates for magnetic field around the astrophysical black holes. 

It is well-known that even small magnetic field $B$ can strongly affect on charged particle's dynamics as a consequence of the large Lorentz force \cite[see, e.g.][]{Jawad16,Hussain17,
DeLaurentis2018PhRvD,Atamurotov13a,Frolov11,Frolov12,
Karas12a,Shaymatov15,
Narzilloev20a,Shaymatov21pdu,Narzilloev20b,Shaymatov21d}. Thus, the magnetic field could be increasingly important to be considered as a background field to test the background geometry in the black hole vicinity. With this motivation there is the family of solution that suggests to involve the interaction between gravity and the axially symmetric magnetic field induced by external sources \cite{Aliev89}. Also another interesting solution that includes the additional gravity that stems from magnetic field describes a static and spherically symmetric black hole with the Melvin’s magnetic universe~\cite{Ernst76}. These kind of solutions are so-called magnetized black holes. In this context the magnetized Reissner-Nordstr\"{o}m black hole solution \cite{Gibbons13} and rotating and charged magnetized black hole solutions with some complicated asymptotic behaviours \cite{Ernst76wz,Aliev89wz,Garcia85wz} have been considered as possible extensions of magnetized black holes. After that the new approach for the magnetized black hole solution was proposed by taking into account the global charge (see for example \cite{Gibbons14wz,Astorino16wz}). Following \cite{Ernst76,Gibbons13} there have been several investigations devoted to the study of the magnetized black hole's properties~\cite{Konoplya08a,Konoplya08b,Shaymatov21c,Shaymatov19b}. 

Recent experiments and modern observations play a decisive role in testing extreme geometric and
remarkable gravitational properties of black holes in general relativity. However, on these lines one can also use astrophysical processes occurring in the environment surrounding astrophysical black holes, i.e. for example X-ray data produced by astrophysical compact objects~\cite{Bambi12a,Bambi16b,Tripathi19} and the quasiperiodic oscillations (QPOs) with the X-ray power observed in microquasars that refer to the low mass X-ray binary systems (i.e. for example a neutron star or a black hole binary systems). {The galactic microquasars are considered as a source of QPOs with the ratio 3/2~\cite{Kluzniak01}.} It is worth noting that QPOs characterized by either low-frequency (LF) or high-frequency
(HF) are observed in the X-ray power spectra. The various kind of HF QPO models was discussed in Refs.~\cite{Stuchlik13A&A,Stella99-qpo,Rezzolla_qpo_03a,Torok05A&A} addressing the epicyclic motion of hot spots and oscillatory models of the accretion disks. It is believed that the above mentioned frequencies characterize the epicyclic motion with corresponding frequencies.  The HF QPOs that will usually exist in
binary system refer to the twin peak HF QPOs that provide information on the matter moving around the compact objects. However, there exists no reasonable model that can explain the appearance of such HF QPOs~\cite{Torok11A&A}. For that reason the epicyclic motion of charged particles around a black hole surrounded by magnetic field has been proposed to explain such phenomenon~\cite{Tursunov20ApJ,Panis19,Shaymatov20egb}. {Also twin peak HF QPOs that usually arise in pairs (i.e., upper $\nu_U$ and lower $\nu_L$ frequencies) have been observed in Galactic microquasars.  For these objects HF QPOs can be observed at the fixed $\nu_U/\nu_L=3/2$ ratio~\cite{Torok05A&A,Remillard06ApJ}.   
The observed upper frequencies, $\nu_U$, are very close to the orbital frequencies of test particles at the stable circular orbit located at the the inner edge of the accretion disc around black holes. Thus, epicyclic motion of test particles with orbital, radial and
latitudinal frequencies can be useful tool in modeling and explaining the observed $\nu_U/\nu_L=3/2$ HF QPOs in the low mass X-ray binary systems. } Following LF and HF QPOs that arise in various parts of the accretion disk there have been several investigations/models addressing the QPOs  \cite[see, e.g.][]{Germana18qpo,Tarnopolski:2021ula,Dokuchaev:2015ghx,Kolos15qpo,Aliev12qpo,Stuchlik07qpo,Titarchuk05qpo,Rayimbaev-Shaymatov21a,Azreg-Ainou20qpo,Jusufi21qpo,Ghasemi-Nodehi20qpo,Rayimbaev22qpo}.

In the present paper we study the dynamics of test particles and epicyclic motions around %a static and spherically symmetric  
the magnetized Ernst black hole. We further aim to constrain the magnetic field with the help of QPOs observed in three microquasars, i.e. GRO J1655-40, XTE J1550-564 and GRS 1915+105. This is what wish to investigate in this paper.  

The paper is organized as follows: In
Sec.~\ref{Sec:metric} we briefly describe the black hole metric. In Sec.~\ref{Sec:motion} we consider the particle dynamics in the environment surrounding the magnetized Ernst black hole. In Sec.~\ref{Sec:qpo} we focus on epecyclic motions with the QPOs in the black hole vicinity and discuss constraints on the magnetic parameter of the magnetized Ernst black hole spcetime in Sec.~\ref{Sec:Constrain}. We present concluding remarks of the obtained results in
Sec.~\ref{Sec:Conclusion}.

Throughout the manuscript we use a system of units in which $G=c=1$.

%\section{Bounds on the energy and angular momentum added}

%\section{Size considerations}
\section{\label{Sec:metric}
Magnetized Ernst Black hole and its electromagnetic field }

\begin{figure*}
  \includegraphics[width=0.4\textwidth]{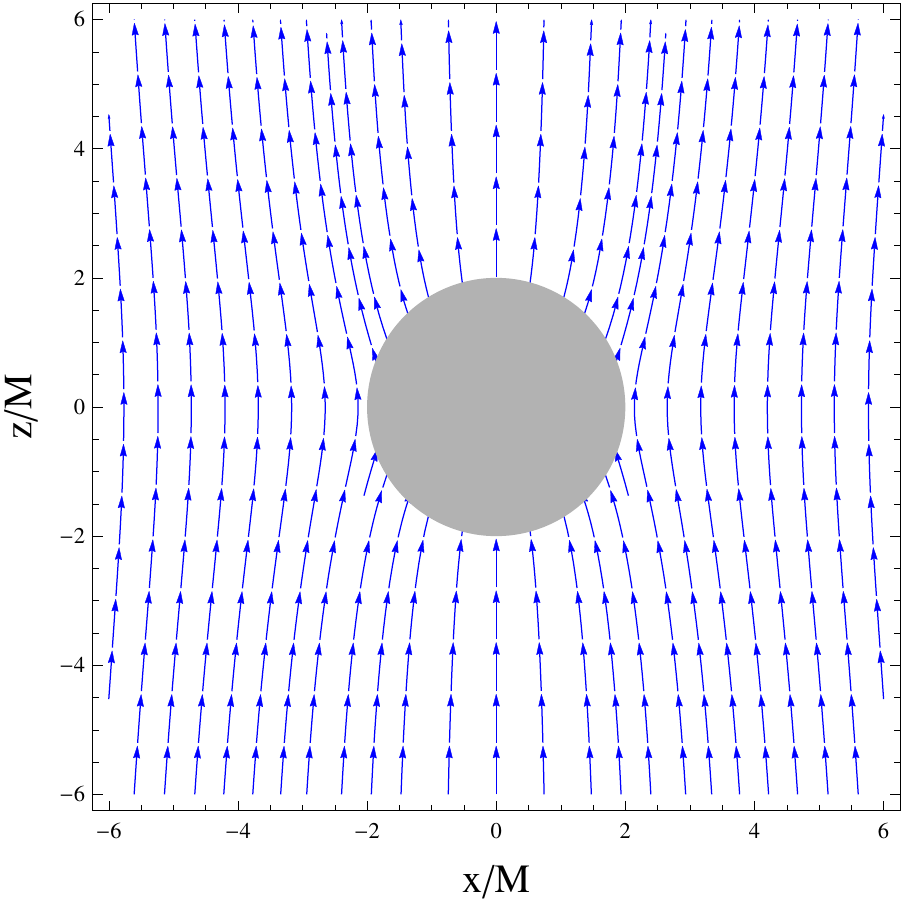}% Here is how to import EPS art
 \includegraphics[width=0.4\textwidth]{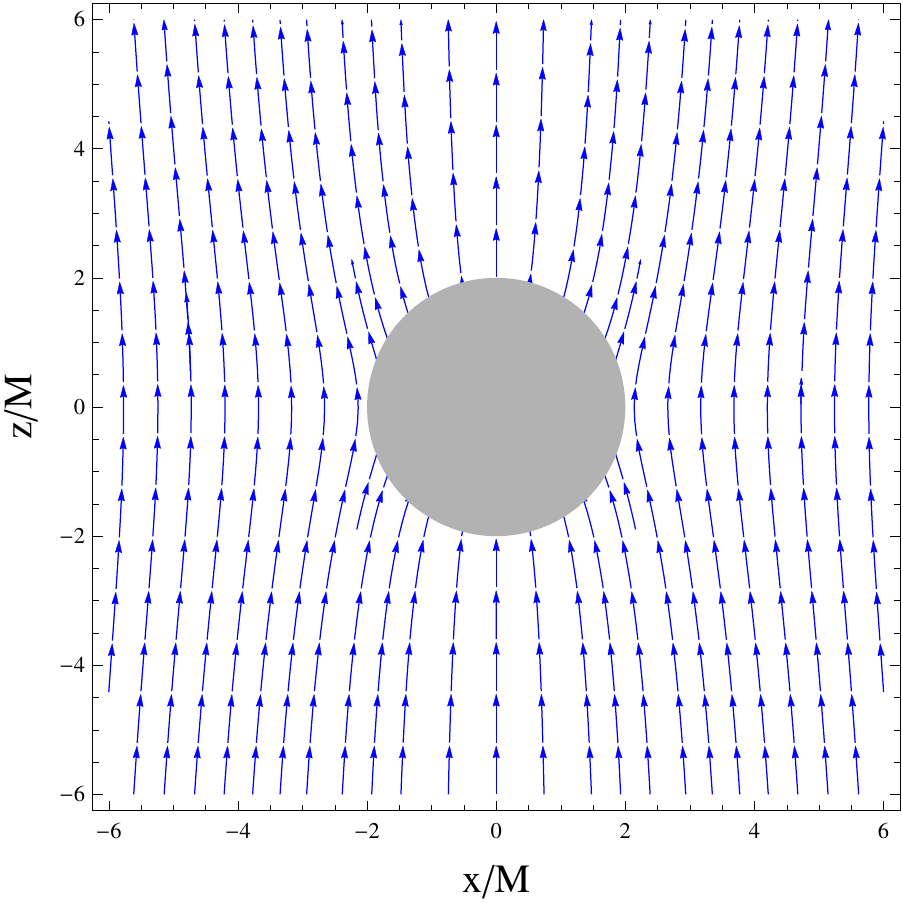}
% c) \includegraphics[width=0.3\textwidth]{con6.eps} %

 \caption{\label{fig:mf} Plot shows the configuration of magnetic field lines in the vicinity of magnetized Ernst black hole. The gray-shaded area shows the black hole horizon. Note that left/right panels respectively refer to $B=0.1/0.5$ {(Note here that we consider $B\to BM$ as a dimensionless quantity, having set $G = c = 1$)}. }
\end{figure*}

Here we briefly review the spacetime metric describing a %static and spherically symmetric 
magnetized Ernst black hole. It is given by 
\begin{align}
ds^2
=\Lambda^2\left(-F(r)dt^2+\frac{dr^2}{F(r)}+r^2d\theta^2\right)+\frac{r^2\sin^2\theta}{\Lambda^2}d\phi^2\,,
\label{eq:metric}
\end{align}
where
\begin{align}
F(r)=& 1-\frac{2M}{r}\,, \\
\Lambda(r,\theta) =& 1+B^2r^2\sin^2\theta\, ,
\end{align}
with the magnetic field parameter $B$. Due to the presence of strong magnetic field the metric is not asymptotically flat and not spherically symmetric either. It is worth noting that the event horizon is given by  $r_{h}=2M$, similarly to what one obtains for the Schwarzschild black hole. The electromagnetic field around the magnetized Ernst black hole has the form as
\begin{align}\label{eq:vec-pot}
A_{\mu}dx^{\mu}=\frac{B r^2 \sin^2\theta}{2\Lambda}d\phi\,.
\end{align}
Since the magnetic field is assumed to be aligned axially, it breaks down the spherical symmetry of the spacetime as well. The underlying geometry of the spacetime is now axis-symmetric as rotations along the $\phi$ direction leave the metric and the electromagnetic field invariant.
The orthonormal components of the magnetic field measured by
zero-angular-momentum observers (ZAMO) with four-velocity
components are given by the following expressions

\begin{eqnarray}
\label{b1}  B^{\hat r}
&&=-\frac{B}{\Lambda}\left(1-\frac{B^2r^2\sin^2\theta}{\Lambda}
\right)\cos\theta \, , \\
\label{b2}  B^{\hat\theta} &&
=\frac{BF(r)^{1/2}}{\Lambda}\left(1-\frac{B^2r^2\sin^2\theta}{\Lambda}
\right)\sin\theta \, .\nonumber\\
\end{eqnarray}
The magnetic field (\ref{b1}) and (\ref{b2}) depends on the
parameter being responsible for the external magnetic field, and
in the limit that \hbox{$M/r\rightarrow 0$} and
\hbox{$\Lambda\rightarrow 1$} we recover the solutions for the
flat spacetime %%
\begin{eqnarray}
 B^{\hat r} =-B\cos\theta,  ~~~ B^{\hat\theta}=B\sin\theta \,,
\end{eqnarray}
which coincides with the homogeneous magnetic field in the Newtonian spacetime as was expected. The configuration of magnetic field lines in the vicinity of magnetized black hole is depicted in Fig~\ref{fig:mf}.

{Let us note that the magnetized Ernst metric is not spherically symmetric, but is actually axi-symmetric due to the fact that it remains invaraint under $\phi=\phi+c$ which is also consistent with the orientation of the electromagnetic field. 
The fact that the spherical symmetry is broken leads to interesting phenomenological aspects such as a possibility to mimic the rotation to some extend. Let us note that one can check the topology of the Ernst solution at the horizon using the Gauss-Bonnet theorem.  At a fixed moment in time $t$, the metric reduces to 
\begin{equation}
ds^2=\Lambda^2\left(\frac{dr^2}{F(r)}+r^2d\theta^2\right)+\frac{r^2\sin\theta^2}{\Lambda^2}d\phi^2\,,
\end{equation}
and using the Gaussian curvature with respect to $g^{(2)}$ on $\mathcal{M}$. Then, the Gauss-Bonnet theorem states that 
\begin{equation}
\iint_{\mathcal{M}} K dA=2 \pi \chi(\mathcal{M}).
\end{equation}
Note that $dA$ is the surface line element of the 2-dimensional surface and $\chi(\mathcal{M})$ is the Euler characteristic number. It is convenient to express sometimes the above theorem in terms of the Ricci scalar, in particular for the 2-dimensional surface at $r=r_h$, then using the Ricci scalar given by
\begin{equation}
\mathcal{R}=\frac{2}{r^2\Lambda^2},
\end{equation}
with $ \sqrt{g^{(2)}}=r^2\sin\theta$, evaluated at $r=r_h$, yielding the following from 
\begin{equation}
\frac{1}{4 \pi}\iint_{\mathcal{M}} \mathcal{R} \sqrt{-g^{(2)}} \,d\theta d\phi=\chi(\mathcal{M}).
\end{equation}
In general the solution of the above integral is complicated due to $\theta$ dependence. For small $B$ we can expand in series and find in leading order terms $\chi=2-8B^2 r_h^2/3$. This shows that the the Euler characteristic number is smaller than $2$, hence in general the topology of Ernst spacetime differs from a perfect sphere.}

%
%%%%%%%%%%%%%%%%%%%%%%%%%%%%%

\section{\label{Sec:motion}
Charged particle dynamics around magnetized Ernst black hole }

Now we focus on charged particle motion around around magnetized Ernst black hole. {We assume that the test particle is endowed with the rest mass $m$ and electric charge $q$. In general, the Hamilton-Jacobi equation of the system is then expressed as
\begin{eqnarray}\label{HJ}
H&=&g^{\mu\nu}\left(\frac{\partial S}{\partial x^\mu}-qA_\mu\right)\left(\frac{\partial S}{\partial x^\nu}-qA_\nu\right)\, ,  
\end{eqnarray}
where $S$ is the action, $x^{\mu}$ is the spacetime coordinates, and $A_\mu$ denotes the vector potential components of the electromagnetic field. 
{Note that the spacetime (\ref{eq:metric}) and the vector potential components (\ref{eq:vec-pot}) are independent of coordinates ($t, \phi$), thus leading to two conserved quantities, i.e. specific energy $E$ and angular momentum $L$. From the properties of the Hamiltonian, it is considered to be as a constant $H=k/2$ with relation to $k/m^2=-1$ for massive particle with $m$ being the mass of particle. For photons one has to set $k=0$. Following to the Hamilton-Jacobi equation, the action $S$ can be written as follows: }
\begin{align}\label{Sol}
S=-\frac{1}{2}k\lambda -Et+L\phi+S_r+S_\theta\, ,
\end{align}
where $S_r$ and $S_\theta$ are the radial and angular functions of $r$ and $\theta$. {Note that $\lambda=\tau/m$ represents the affine parameter with proper time $\tau$}. Inserting equation (\ref{Sol}) into (\ref{HJ}) the Hamilton equations reads
\begin{eqnarray}\label{Eq:separable}
&k=& -\frac{F(r)^{-1}}{\Lambda^2}E^2
+ \frac{F(r)}{\Lambda^2 }
\left(\frac{\partial S_{r}}{\partial r}\right)^2
+\frac{1}{\Lambda^2 r^2}\left(\frac{\partial S_{\theta}}{\partial \theta}\right)^2\nonumber\\&&+
\frac{\Lambda^2\left(L-qA_{\varphi}\right)^2}{r^2\sin^2\theta} \, . 
\end{eqnarray}
{The Hamiltonian can be separated into dynamical and potential parts, i.e. $H=H_{\textrm{dyn}}+H_{\textrm{pot}}$ with } 
\begin{eqnarray}
%H_{\textrm{dyn}}&=& \frac{1}{2}\left[\frac{F(r)}{\Lambda^2} \left(\frac{\partial S_{r}}{\partial r}\right)^2 +\frac{1}{\Lambda^2 r^2}\left(\frac{\partial S_{\theta}}{\partial \theta}\right)^2\right],\\
H_{\textrm{dyn}}&=& \frac{1}{2}\left[\frac{1}{g_{rr}} \left(\frac{\partial S_{r}}{\partial r}\right)^2 +\frac{1}{g_{\theta\theta}}\left(\frac{\partial S_{\theta}}{\partial \theta}\right)^2\right],\\
H_{\textrm{pot}}&=& \frac{1}{2}\left[\frac{\mathcal{E}^2}{g_{tt}}+\frac{\left(\mathcal{L}-\frac{q}{m}A_{\varphi}\right)^2}{g_{\phi\phi}}+1\right]\ .\label{Eq:sepham}
\end{eqnarray}}
{Here we note that we further use the potential part of the Hamiltonian, $H_{\textrm{pot}}$, to define the epicyclic frequencies.}

{For further analysis we shall restrict motion for charged test particle to the equatorial plane (i.e., $\theta=\pi/2$).} Following Eq.~(\ref{Eq:separable}), we obtain the radial equation of motion for charged particles in the following form
\begin{eqnarray}\label{Eq:rdot}
\dot{r}^2=\Big(\mathcal{E}-\mathcal{E}_-(r)\Big)\Big(\mathcal{E}-\mathcal{E}_+(r)\Big)\, ,
\end{eqnarray}
{where $\mathcal{E}_{+}(r)$ and $\mathcal{E}_{-}(r)$ are the two roots of the equation $\dot{r}=0$ %refer describes the radial function of the radial motion
}.  As seen from Eq.~(\ref{Eq:rdot}),  we have either $\mathcal{E}>\mathcal{E}_{+}(r)$ or $\mathcal{E}<\mathcal{E}_{-}(r)$ since $\dot{r}^2\geq 0$. {However, we shall restrict ourselves to the case $\mathcal{E}_{+}(r)$ which is physically acceptable and consequently we select $\mathcal{E}_{+}(r)$ as a effective potential,  i.e. $V_{\rm eff}(r)=\mathcal{E}_{+}(r)$.} Thus we have    
\begin{eqnarray} \label{Veff}
V_{\rm eff}(r)= \left(1-\frac{2M}{r}\right)^{1/2}\left[{\Lambda^2}
+\frac{\Lambda^4}{r^2}\left(\mathcal{L}-\frac{q\,B\,r^2}{2m\Lambda}\right)^2\right]^{1/2} \, . \nonumber\\
\end{eqnarray}
{Here, we denote parameters $\mathcal{E}=E/m$ and  $\mathcal{L}=L/(M m)$.} 
From Eq.~(\ref{Veff}), the effective potential can easily recover the one as in the Schwarzschild spacetime in case we eliminate all parameters except black hole mass $M$.   
As can be seen from Eq.~(\ref{Veff}), the vector potential is involved in the expression of the effective potential due to the fact that charged particle depends upon the magnetic field component. 

\begin{figure*}
\centering
\includegraphics[width=0.3\textwidth]{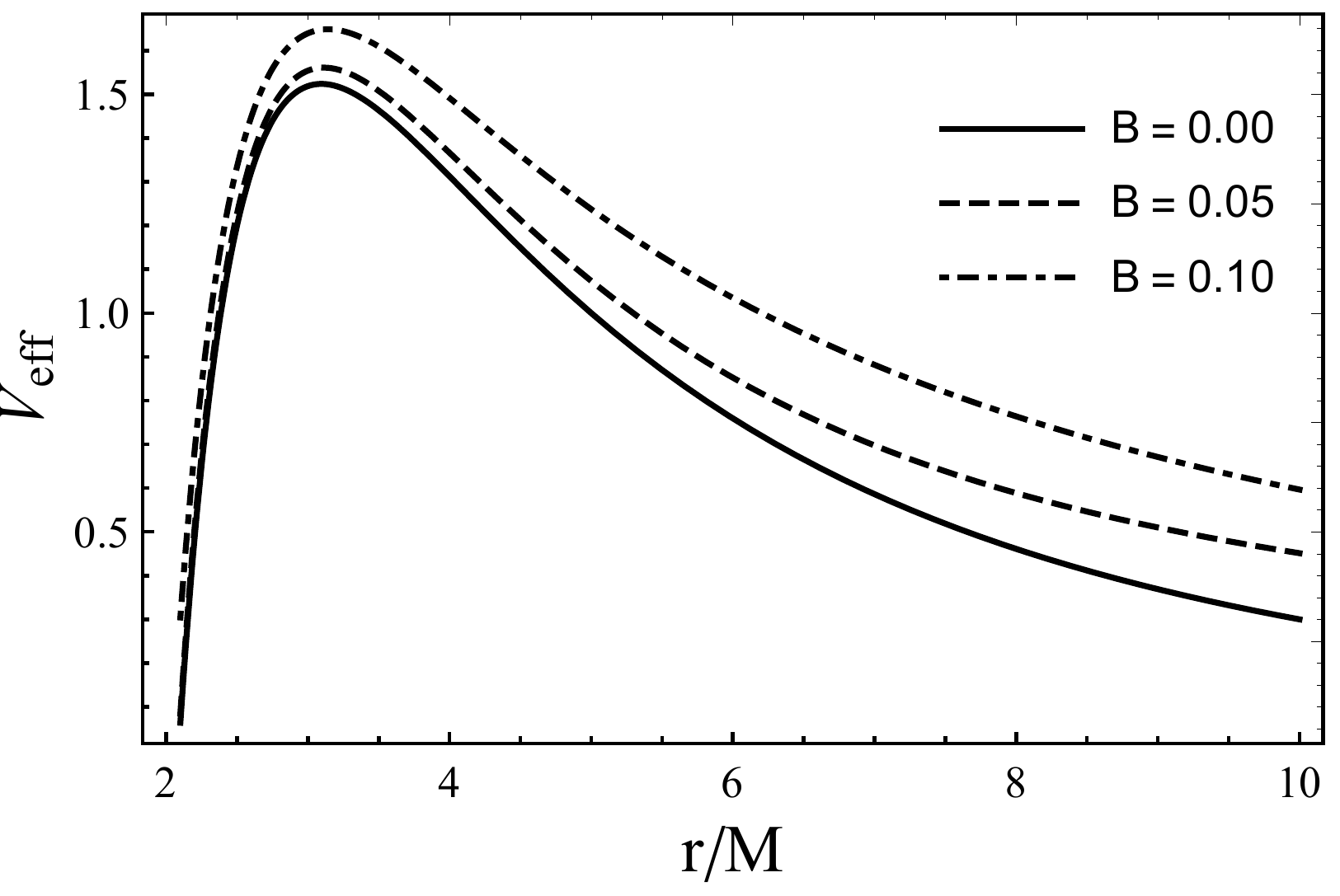}% Here is how to import EPS art
\includegraphics[width=0.3\textwidth]{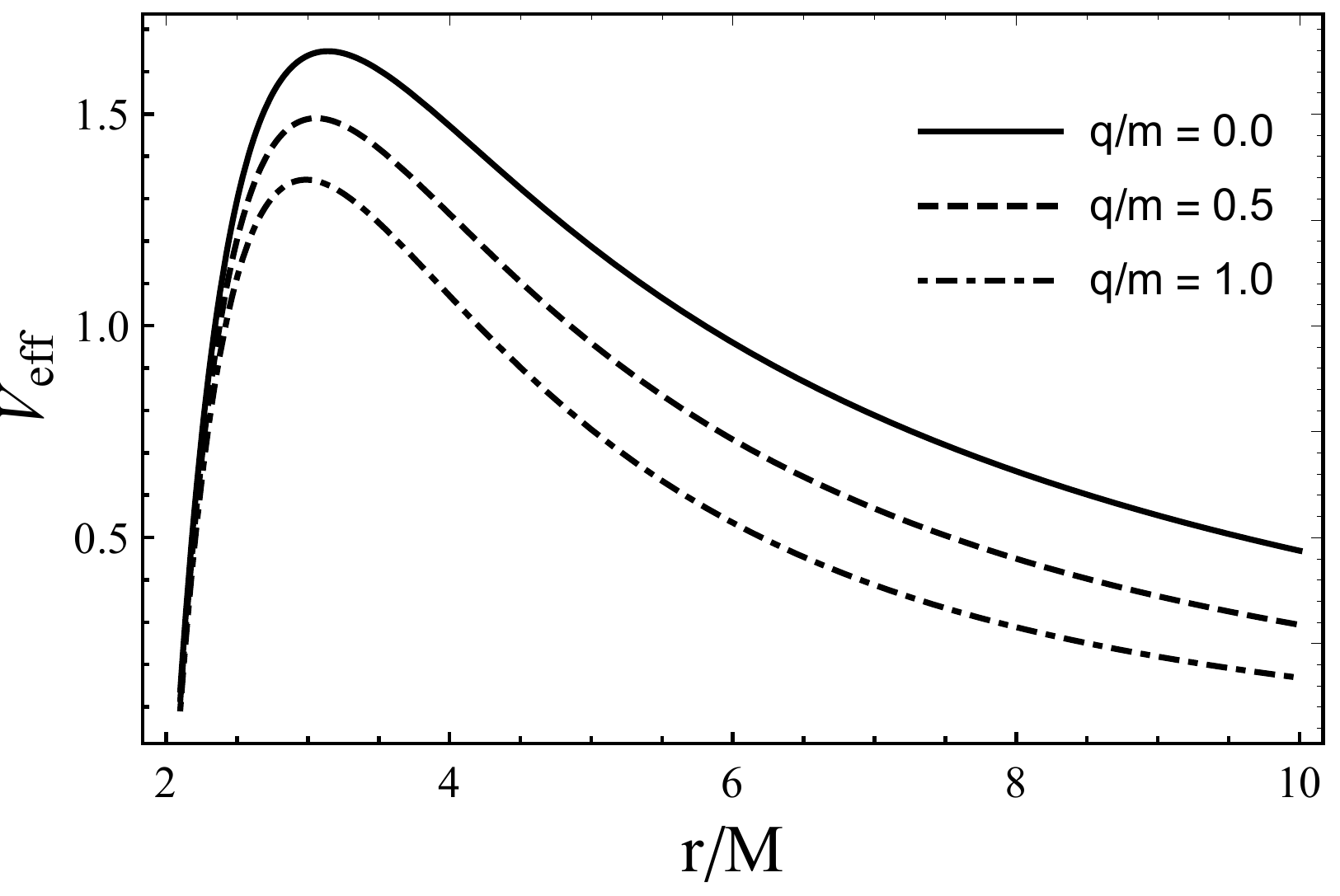}
\includegraphics[width=0.3\textwidth]{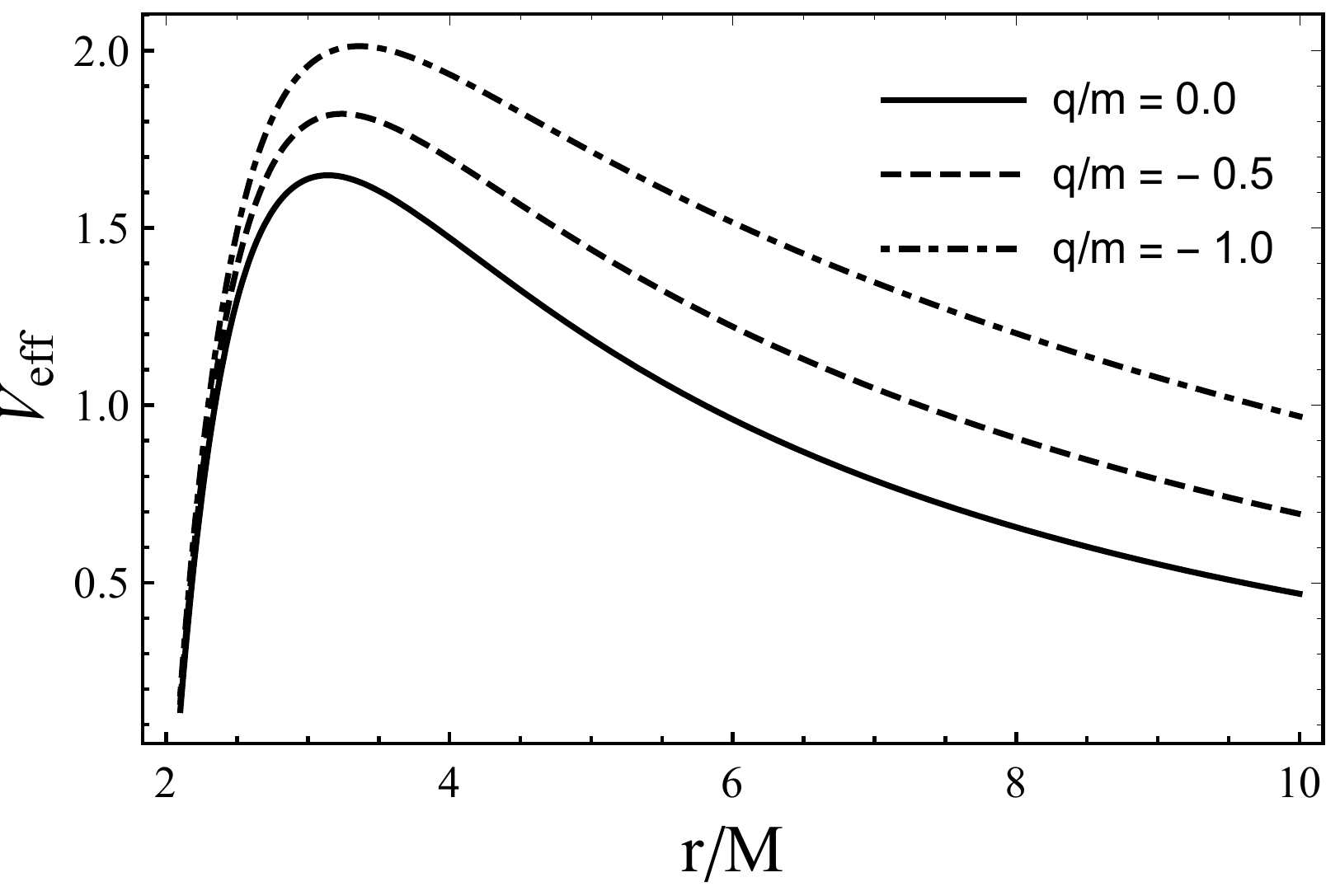}% Here is how to import EPS art
%f)  \includegraphics[width=0.45\textwidth]{6.eps}
%
\caption{\label{feff} Radial dependence of the effective potential for the radial motion of test particles around the magnetized Ernst black hole. {Left panel: $V_{\rm eff}$ is plotted for various combination of $B$ for neutral particle  case, i.e. $q/m=0$. Middle/right panels: $V_{\rm eff}$ is plotted for various combinations of charged particle parameter $q/m$ for the fixed magnetic field parameter $B=0.1$.}}
\end{figure*}
\begin{figure*}
\centering
\includegraphics[width=0.3\textwidth]{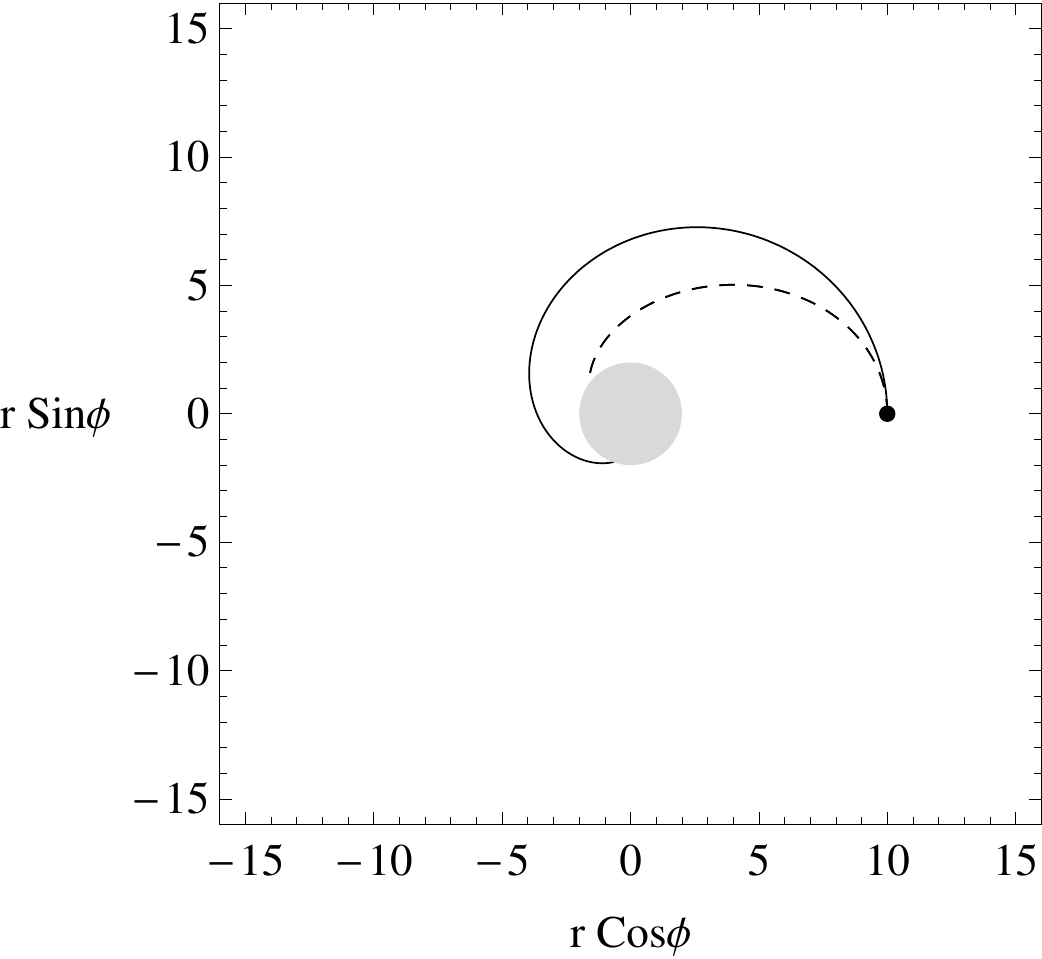}% Here is how to import EPS art
\includegraphics[width=0.3\textwidth]{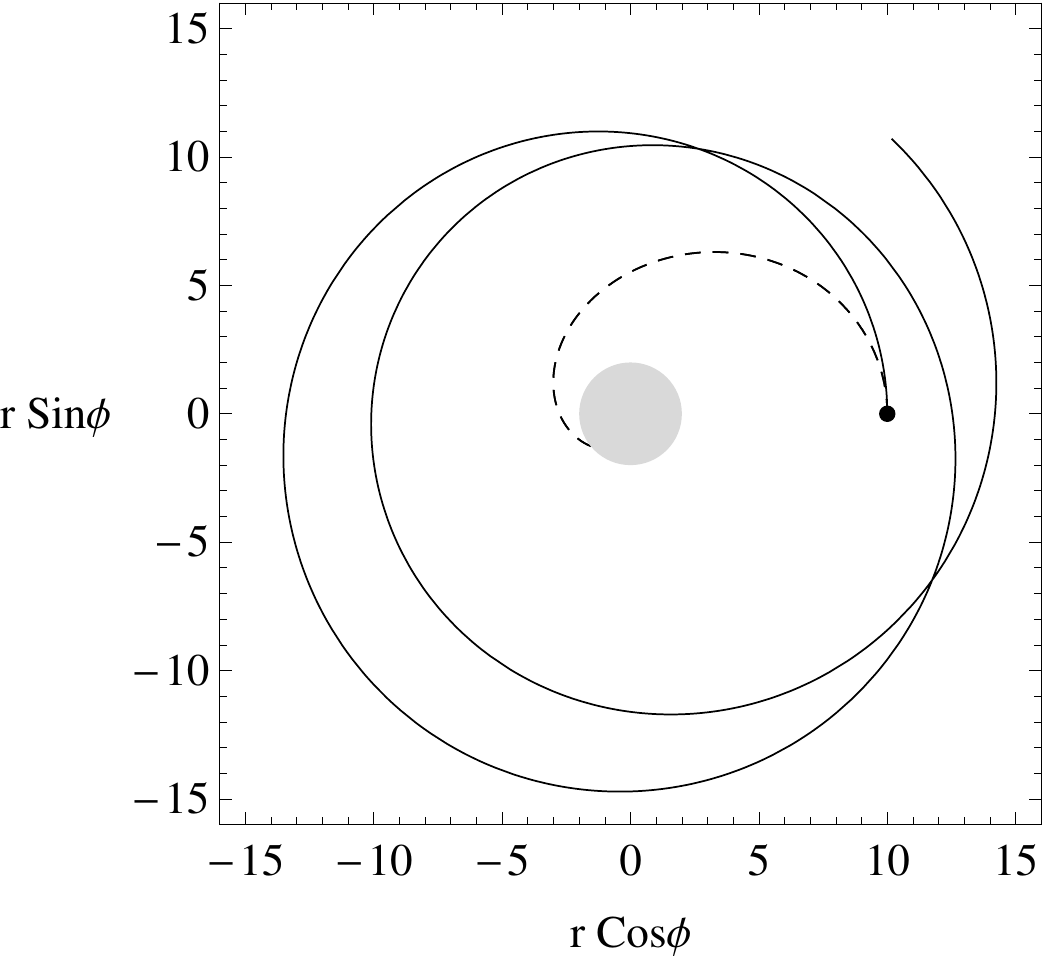}
\includegraphics[width=0.3\textwidth]{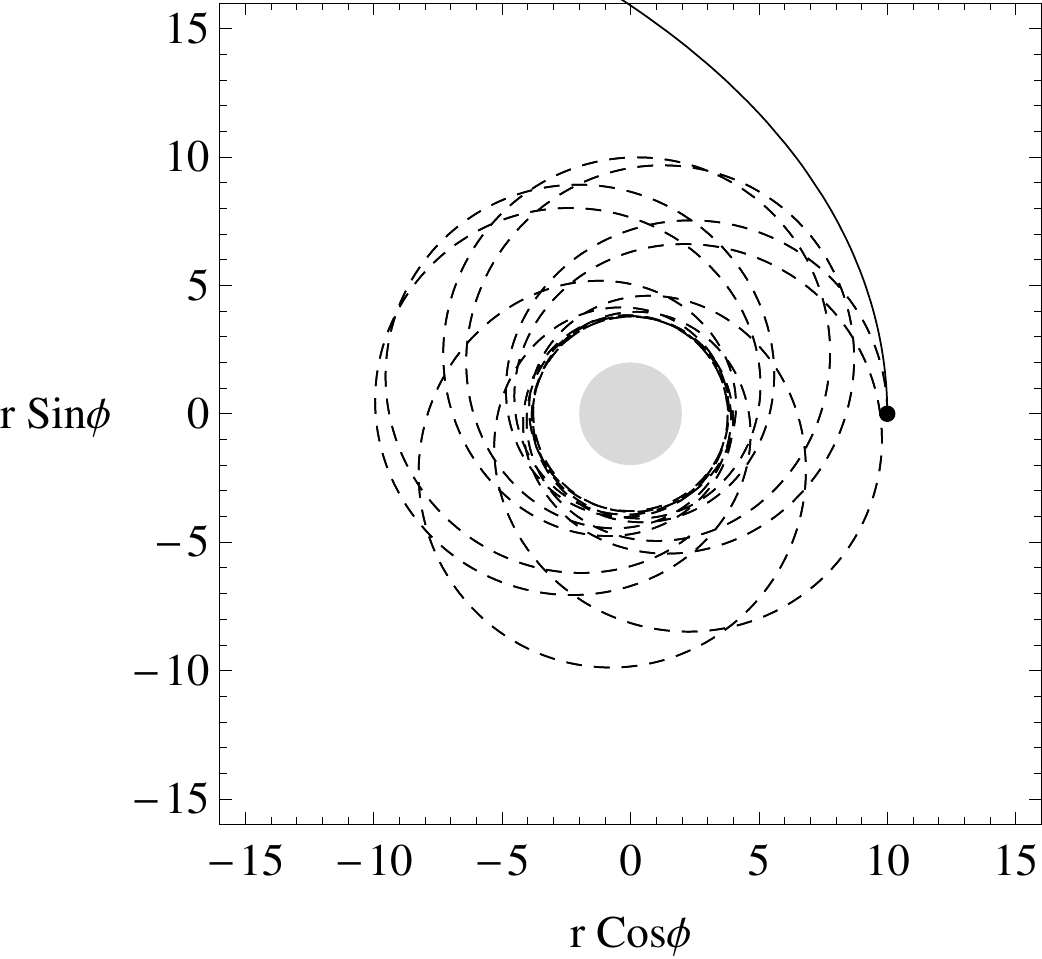}% Here is how to import EPS art

\caption{\label{trajectory}  Trajectories of the particle at the
equatorial plane around the magnetized Ernst black hole for magnetic field parameter $B=0$ (solid curve) and
$B=0.04$ (dashed curve) for various combination of angular momentum $\mathcal{L}=3$ (left panel), $\mathcal{L}=4$ (middle panel), and $\mathcal{L}=6$ (right panel) for all possible orbits. {Note that particle starts from $r_{0}/M=10$ towards the black hole.}}
\end{figure*}
%
%\subsubsection{Effective potential}

Let us then analyze the effective potential in order to understand more deeply the radial motion of test particles moving around the black hole. In Fig.~\ref{feff}, we show the radial
dependence of $V_{\rm eff}$ radial motion of neutral/charged particle.
In Fig.~\ref{feff}, left panel demonstrates the impact of the magnetic field parameter $B$ on the profile of the effective potential for neutral particle, while middle and right panels demonstrate the impact of parameter $q/m$ for charged particle. As can be seen from left panel of Fig.~\ref{feff}, the shape of the effective potential shifts upward as a consequence of an increase in the value of magnetic field parameter $B$, thus strengthening gravitational potential. Similarly, for negative values of charged particle parameter $q/m$ right panel illustrates the same behaviour as the one for left panel. However, the shape of the effective potential shifts downward with increasing positive values of $q/m$, thereby giving rise to the decrease in the strength of gravitational potential, as seen in Fig.~\ref{feff}. Also note that as a consequence of positive charged particle stable circular orbits shift to left to smaller $r$. 
\begin{figure*}
\begin{tabular}{c c }
  \includegraphics[scale=0.55]{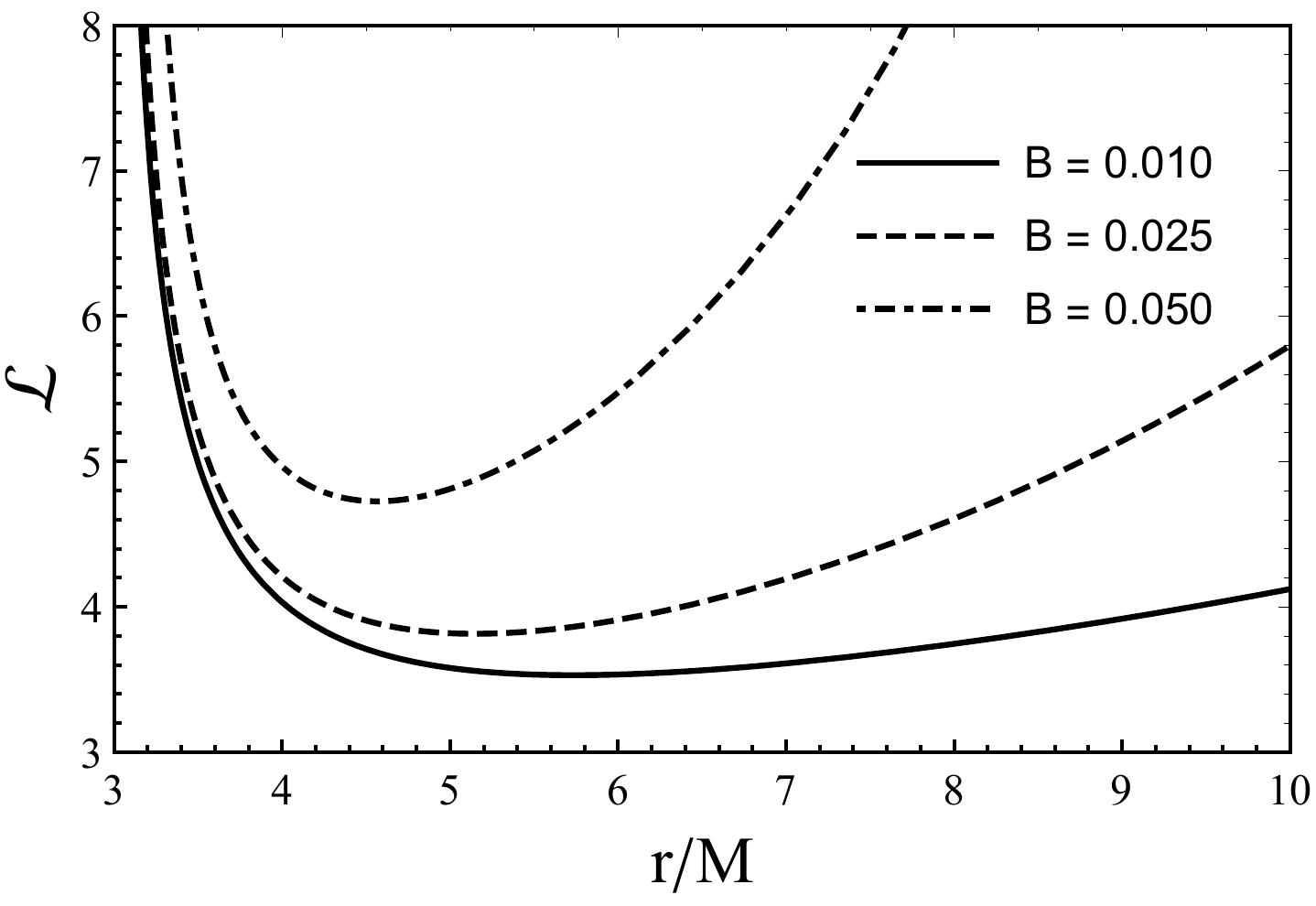}\hspace{-0.0cm}
  &  \includegraphics[scale=0.55]{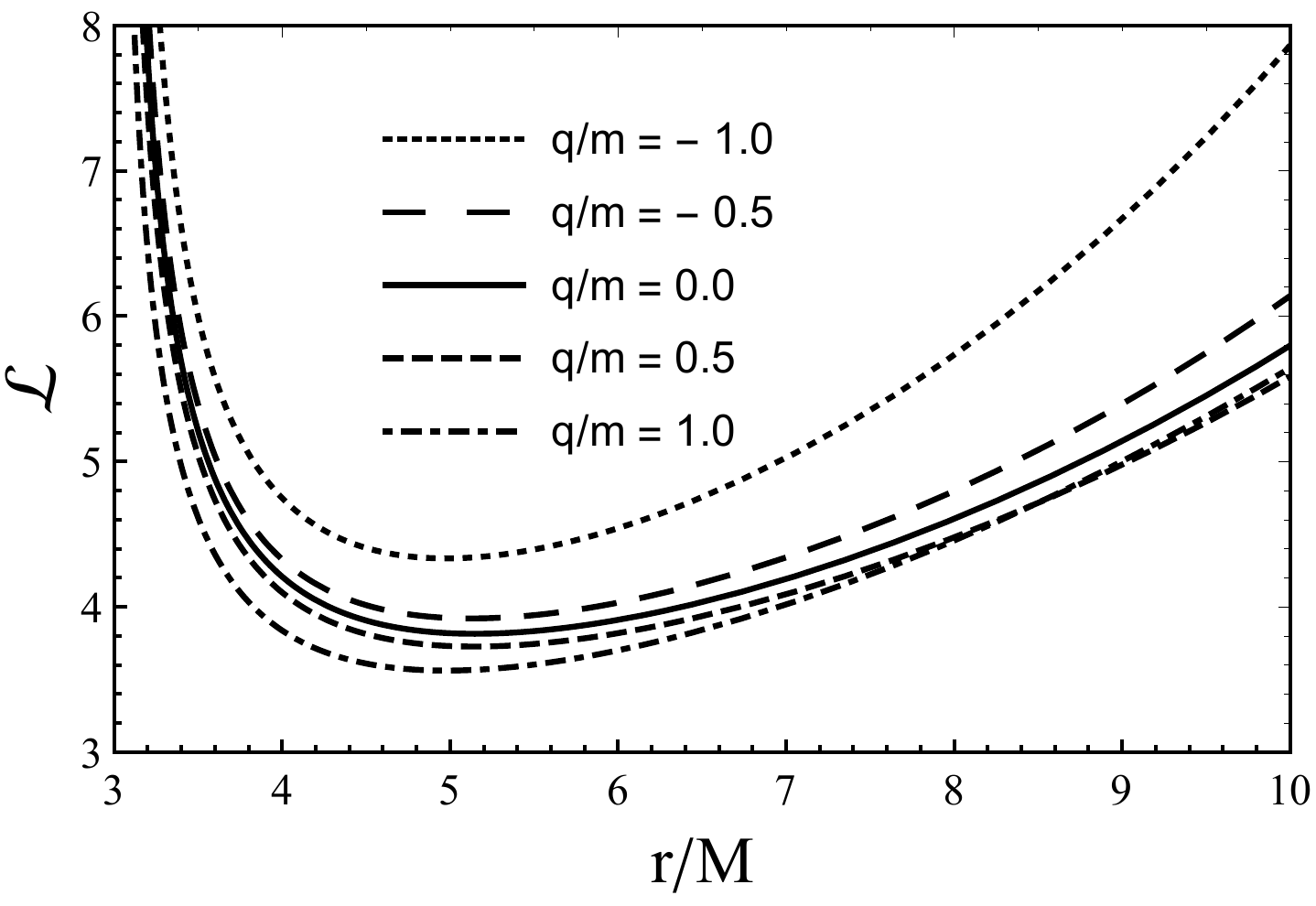}
\end{tabular}
\caption{\label{fig:ang} Plot shows the radial dependence of the angular momentum as the function of $r/M$ in the equatorial plane (i.e. $\theta=\pi/2$). Left panel: $\mathcal{L}$ is plotted for various combinations of magnetic field parameter $B$ for neutral particle case, i.e. $q/m=0$. Right panel: $\mathcal{L}$ is plotted for various combinations of charged particle parameter $q/m$ for the fixed $B=0.025$. }
\end{figure*}

%\subsubsection{Trajectory of the particles around Ernst black hole }
We also analyze the particle trajectories of test particles moving around the magnetized Ernst black hole. Here, we demonstrate the trajectory of particles at the the equatorial plane. In Fig.~\ref{trajectory}, all plots show various behavior of the particle trajectory around the magnetized Ernst black hole. It is increasingly important to understand more deeply the behaviour of possible orbits and trajectories of particles around the black hole. With this in view, we consider the terminating orbits (left panel), the bound orbits and the escape orbits for particles. As can be seen in Fig.~\ref{trajectory}, middle panel shows the bound orbits that appear in the balance between the centrifugal force and the gravitational force that stems from the parameters $B$ and $M$, whereas in right panel there appear no bound orbits as the particle can escape from the pull of the black hole when the centrifugal force dominates over the gravitational one as a consequence of absence of magnetic field parameter $B=0$. Thus, in this case the overall force becomes repulsive. It becomes however attractive once the magnetic field parameter is involved, and thus particle orbits become more unstable that allow particles to fall into
the black hole, as seen in Fig.~\ref{trajectory}. 
\begin{figure}[ht]
%  \centering
\includegraphics[scale=0.52]{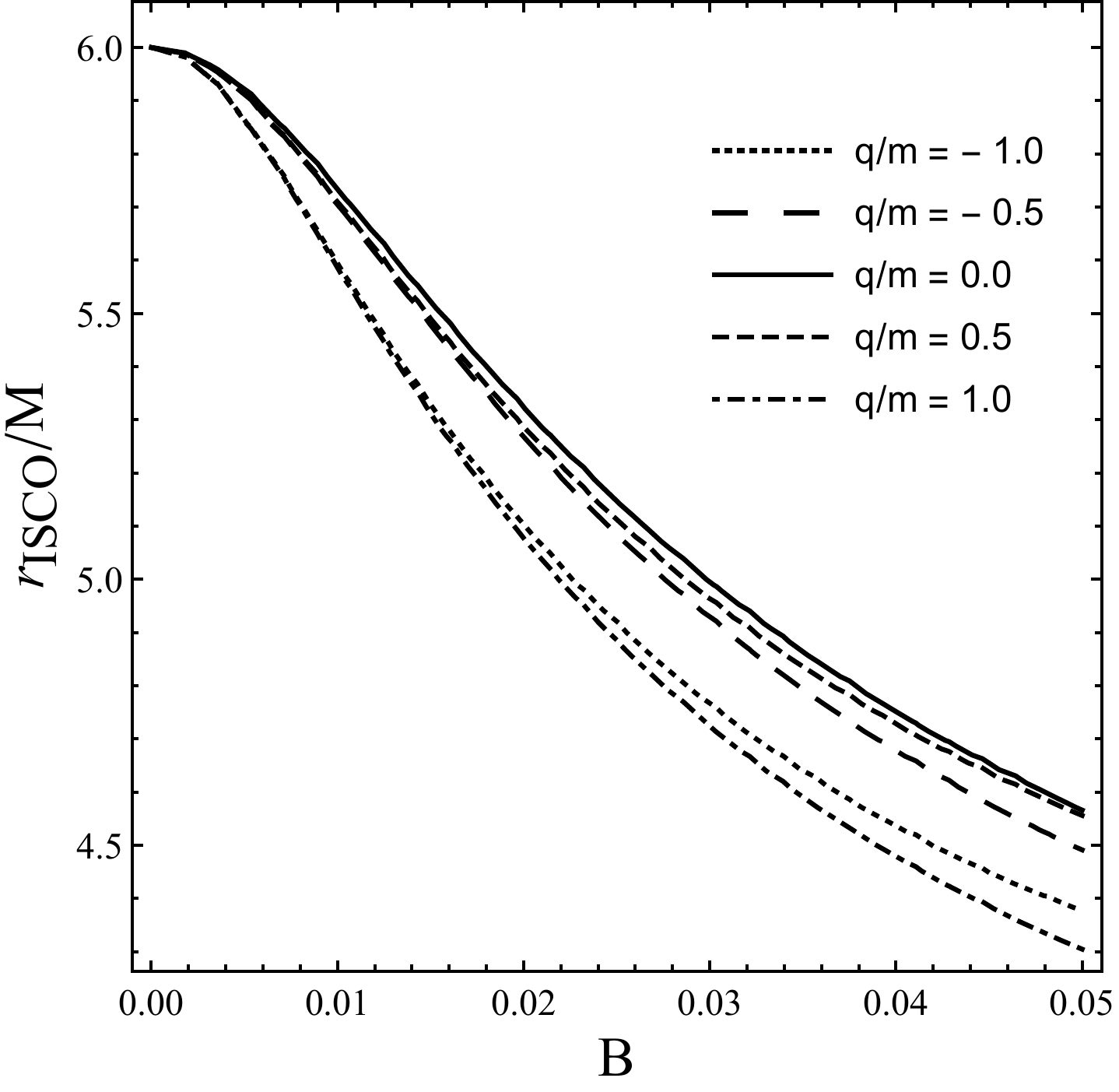}
\caption{\label{fig:isco} Plot shows the ISCO radius as the function of $B$ in the case with both neutral and charged particles. }
\end{figure}

Let us then consider stable circular orbits around the magnetized Ernst black hole. For particles to be at circular orbits one needs to solve the following equations simultaneously
\begin{eqnarray}\label{Eq:cir1}
V_{\rm eff}(r)=\mathcal{E}\, \mbox{~~and~~} V_{\rm eff}^{\prime}(r)=0\, ,
\label{Eq:cir2}
\end{eqnarray}
where $^{\prime}$ refers to a derivative with respect to $r$. For particles to be at the circular orbits one needs to have the following angular momentum $\mathcal{L}$  
\begin{eqnarray}\label{Eq:ang_circular}
\mathcal{L}_{\pm}&=& \frac{\Lambda^{-2} }{2 \big(r \Lambda F(r)'-2 F(r) \left(\Lambda-2 r \Lambda'\right)\big)}\nonumber\\&\times &\Big[(q/m)\, B r^3 \Lambda \left(\Lambda F(r)'+3 F(r) \Lambda'\right)\nonumber\\&\pm &\Big(r^3 \Lambda^2 \left(F(r)^2 \left(-4 \left((q/m)^2\,B^2 r^2-4\right) \Lambda \Lambda'\right.\right. \nonumber\\&+& \left. r \left((q/m)^2\,B^2 r^2-32\right) \Lambda'^2+4 (q/m)^2\,B^2 r \Lambda^2\right)\nonumber\\&+& \left. 8 F(r) \Lambda F(r)' \left(\Lambda-3 r \Lambda'\right)-4 r \Lambda^2 F(r)'^2\right)\Big)^{1/2}\Big]\, .\nonumber\\
\end{eqnarray}
%\end{widetext}
%
%
In Fig.~\ref{fig:ang} we show the radial dependence of $\mathcal{L}$ {(Note that through by $\mathcal{L}$ we
would mean positive $\mathcal{L}_{+}$)} of test particles circular orbits around the magnetized Ernst black hole for various combinations of  magnetic field and charged particle parameters. We note that as a consequence of the presence of magnetic field parameter $B$ {circular orbits shift towards left to smaller $r$}, thus increasing the value of $\mathcal{L}$ for particles to be on circular orbits with small the radii (see Fig.~\ref{fig:ang}, left panel). However, we show the dependence on $\pm q/m$  of the angular momentum for charged particles in circular orbits (see Fig.~\ref{fig:ang}, right panel). 
{Also, it is clearly seen from Fig.~\ref{fig:ang} that} both $-q/m$ and $B$ have similar effect, thereby reducing the radii of circular orbits. 

We shall now determine the innermost stable circular orbit (ISCO) for test particles orbiting around the magnetized Ernst black hole. The radii of ISCO is defined by second derivative of $V_{\rm eff}$ as follows: 
\begin{eqnarray}\label{Eq:isco}
V_{\rm eff}^{\prime\prime}(r)=0\, .
\end{eqnarray}
From the last equation, we determine the ISCO radius using Eq.~(\ref{Eq:ang_circular}).  
In Fig.~\ref{fig:isco}, we show the dependence on magnetic field parameter $B$ of $r_{ISCO}$ for the test particles. As can be seen from Fig.~\ref{fig:isco}, the ISCO radius gets decreased as a consequence of an increase in the value of $B$ for all cases. However, one can notice that $r_{ISCO}$ is slightly influenced due to the presence of opposite charged particles, $\pm q/m$, while keeping the magnetic field parameter $B$ fixed.

\section{\label{Sec:qpo}
Epicyclic frequencies }

We now consider the periodic motion of the test particle which orbits at the stable circular orbits. For that all orbits that the particle can move on should occur $r\geq r_{isco}$ determined by the minimum of $V_{\rm eff}(r)$. The given a small perturbation $r = r_0 + \delta r$ and $\theta=\pi/2+\delta \theta$, the particle starts to oscillate around the circular orbit with $r_0$ with so-called radial and latitudinal frequencies that respectively refer to the so-called epicyclic motion. Thus, for the given small perturbation $\delta r$ and $\delta \theta$ one can write the equations for a linear harmonic oscillation as follows:
\begin{eqnarray}
\label{Eq:har_osc}
\delta\ddot{ r}+\bar{\omega}_r^2 \delta r = 0, \ \ \ 
\delta\ddot{ \theta}+\bar{\omega}_\theta^2 \delta\theta = 0,
\end{eqnarray}
with the radial $\bar{\omega}_r$ and the latitudinal $\bar{\omega}_\theta$ frequencies for epicyclic oscillations. Note that the dot in the above equation denotes derivative with respect to the proper time. These frequencies are measured by a local observer and defined by the following equations~\cite{Shaymatov20egb,Stuchlik21_qpo}
\begin{eqnarray}
\label{baromegas}
\bar{\omega}_r^2 &=& \frac{1}{g_{rr}}\frac{\partial^2 H_{\textrm{pot}}}{\partial r^2} ,\\ 
\bar{\omega}_\theta^2 &=& \frac{1}{g_{\theta\theta}}\frac{\partial^2 H_{\textrm{pot}}}{\partial \theta^2}, \\ 
\bar{\omega}_\phi &=& \frac{1}{g_{\phi\phi}}\Big(\mathcal{L}-\frac{q}{m} A_\phi\Big)\, .
\end{eqnarray}
Before representing the above epicyclic frequencies $\omega_{r,\theta}$ %From the above equations 
we now consider the periodic motion for both neutral and charged particles. For that we write the normalization condition, $u^{\alpha}u_{\alpha}=-1$, for particles in the following way 
\begin{eqnarray}
g_{tt}(u^{t})^2+g_{rr}(u^{r})^2+g_{\theta\theta}(u^{\theta})^2+g_{\phi\phi}(u^{\phi})^2=-1\, ,
\end{eqnarray}
For the periodic motion for the rest particle that orbits around a black hole with the fundamental frequencies , i.e. Keplerian and Larmor frequencies, one needs to focus on the stable circular orbits for which $u^{\alpha}= (u^{t}, 0, 0,u^{\phi})$. With this in view, we have following equations 
\begin{eqnarray}\label{Eq:ut}
u^{t}&=&\frac{1}{\sqrt{-g_{tt}-\omega^2g_{\phi\phi}}}\, ,\\
\label{Eq:en}
\mathcal{E}&=&-\frac{g_{tt}}{\sqrt{-g_{tt}-\omega^2g_{\phi\phi}}}\, ,\\
\label{Eq:ang}
\mathcal{L}&=&\frac{g_{\phi\phi}\omega}{\sqrt{-g_{tt}-\omega^2g_{\phi\phi}}}+\frac{q}{m}A_{\phi}\, ,
\end{eqnarray}
where we have defined $\omega= \frac{d\phi}{dt}$.  
We then obtain general expression for the orbital frequency $\omega$ referred to as so-called Keplerian frequency using the non-geodesic equation for charged particles  
\begin{eqnarray}
g_{tt,r}+\omega^2g_{\phi\phi,r}&=&-\frac{2q}{m}\big({A_{t,r}}+\omega A_{\phi,r}\big)\nonumber\\&\times &\left(-g_{tt}-\omega^2g_{\phi\phi}\right)^{1/2}\, .
\end{eqnarray}
For the Keplerian frequency, $\omega_k$, the above equation solves to give 
\begin{eqnarray}\label{Eq:kep}
\omega^2_{k}&=&\Bigg[{\omega_{0}^2}-2{g_{tt}}\left(\frac{qA_{\phi,r}}{mg_{\phi\phi,r}}\right)^2\pm %2\omega_{0}^2\frac{qA_{\phi,r}}{mg_{\phi\phi,r}}
\frac{2qA_{\phi,r}}{mg_{\phi\phi,r}} \nonumber\\&\times&\left\{-\omega_{0}^2g_{tt} -\omega_{0}^4g_{\phi\phi}+\left(\frac{qA_{\phi,r}\,g_{tt}}{mg_{\phi\phi,r}}\right)^2 \right\}^{1/2}\Bigg]\nonumber\\&\times& \left[1+4g_{\phi\phi}\left(\frac{qA_{\phi,r}}{mg_{\phi\phi,r}}\right)^2\right]^{-1}\, .
\end{eqnarray}
This clearly shows that as a consequence of $q=0$ the above equation turns simple and is given by $\omega_{0}^2={-g_{tt,r}/g_{\phi\phi,r}}$ that referes to the Keplerian frequency for neutral particle orbiting around the magnetized black hole. Note that for $q\neq0$ Eq.~(\ref{Eq:kep}) turns out to be very long and complicated expression for explicit display. We therefore further resort to numerical evolution for $\omega_k$. However, in the case of neutral particle $q=0$ Eq.~(\ref{Eq:kep}) has the form as
\begin{eqnarray}
\omega_{k}=\sqrt{\frac{\left(1+B^2 r^2\right)^4 \left(M+2 B^2 r^3-3 M B^2 r^2\right)}{r^3 \left(1-B^2 r^2\right)}}\, .
\end{eqnarray}

Using Eqs.~(\ref{Eq:ut}-\ref{Eq:ang}) and (\ref{Eq:kep}) one can define epicyclic frequencies $\omega_{r}$ and $\omega_{\theta}$ as follows: 
\begin{eqnarray}\label{Eq:wr}
\bar{\omega}_{r}^2&=& \left[-\frac{g_{tt,r}^2}{g_{tt}g_{rr}}+\frac{g_{tt,rr}}{g_{rr}}-\frac{\omega_{k}^2}{g_{rr}}\left(\frac{g_{\phi\phi,r}^2}{g_{\phi\phi}}-\frac{g_{\phi\phi,rr}}{2}\right)\right. \nonumber\\&+&
\frac{q}{m}\frac{\omega_{k}}{g_{rr}}\left(A_{\phi,rr}-2A_{\phi,r}\frac{g_{\phi\phi,r}}{g_{\phi\phi}}\right)\left(-g_{tt}-\omega_{k}^{2}g_{\phi\phi}\right)^{1/2}\nonumber\\&+&\left.\left(\frac{q}{m}\right)^2\frac{A_{\phi,r}^2}{g_{rr}}\left( \omega_{k}^2+\frac{g_{tt}}{g_{\phi\phi}}\right)\right]\left(\frac{1}{g_{tt}+\omega_{k}^{2}g_{\phi\phi}}\right)\, ,\\ \nonumber\\
\label{Eq:wth}
\bar{\omega}_{\theta}^2&=& \left[-\frac{g_{tt,\theta}^2}{g_{tt}g_{\theta\theta}}+\frac{g_{tt,\theta\theta}}{g_{\theta\theta}}-\frac{\omega_{k}^2}{g_{\theta\theta}}\left(\frac{g_{\phi\phi,\theta}^2}{g_{\phi\phi}}-\frac{g_{\phi\phi,\theta\theta}}{2}\right)\right. \nonumber\\&+&
\frac{q}{m}\frac{\omega_{k}}{g_{\theta\theta}}\left(A_{\phi,\theta\theta}-2A_{\phi,\theta}\frac{g_{\phi\phi,\theta}}{g_{\phi\phi}}\right)\left(-g_{tt}-\omega_{k}^{2}g_{\phi\phi}\right)^{1/2}\nonumber\\&+&\left.\left(\frac{q}{m}\right)^2\frac{A_{\phi,\theta}^2}{g_{\theta\theta}}\left( \omega_{k}^2+\frac{g_{tt}}{g_{\phi\phi}}\right)\right]\left(\frac{1}{g_{tt}+\omega_{k}^{2}g_{\phi\phi}}\right)\, .
\end{eqnarray}
From the above equations the epicyclic frequencies {turn out to be complicated expression for explicit display}. Thus we further give numerical analysis to represent them explicitly.
\begin{figure}
\begin{tabular}{c }
  \includegraphics[scale=0.6]{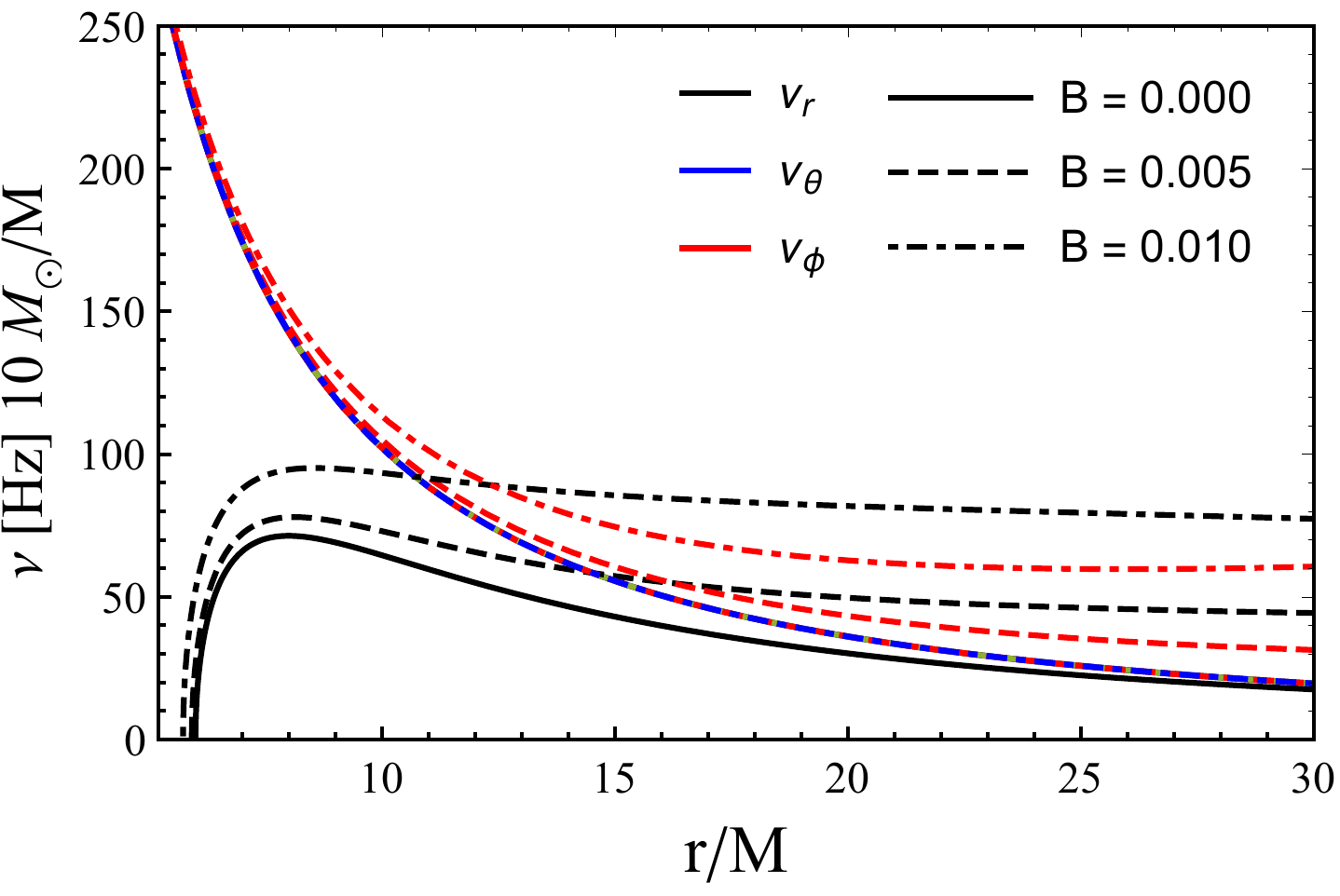}\hspace{-0.0cm}
  
\end{tabular}
\caption{\label{fig:qpo_freq} Plot shows the epicyclic frequencies as a function of $r/M$ in the case with neutral particle. Radial, latitudinal and orbital frequencies are plotted for various combinations of magnetic field parameter $B$. Note that solid lines refer to the epicyclic frequencies for the Schwarzschild black hole. }
\end{figure}

\begin{figure*}
\begin{tabular}{c c}
  \includegraphics[scale=0.55]{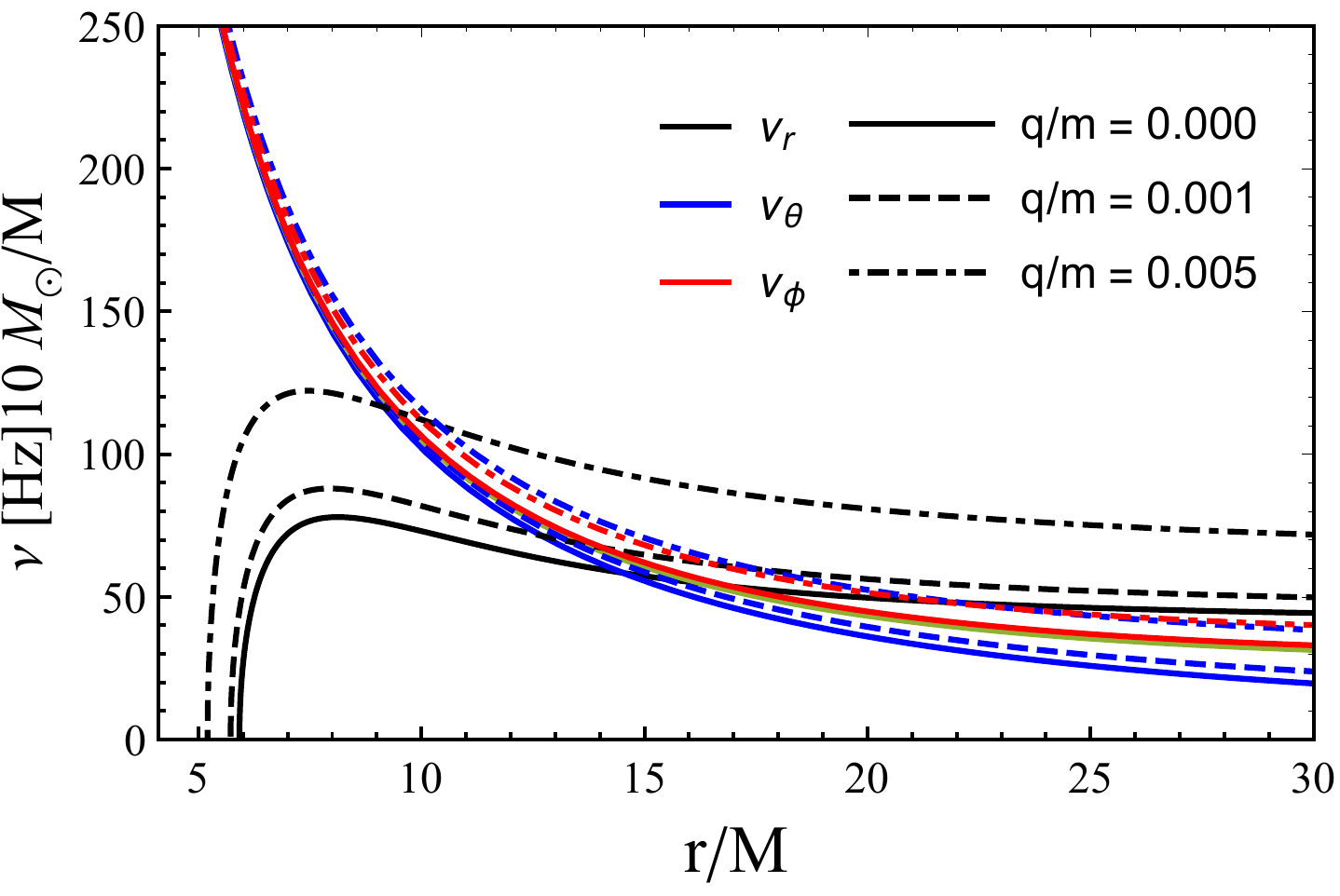}\hspace{-0.0cm}
   &  \includegraphics[scale=0.55]{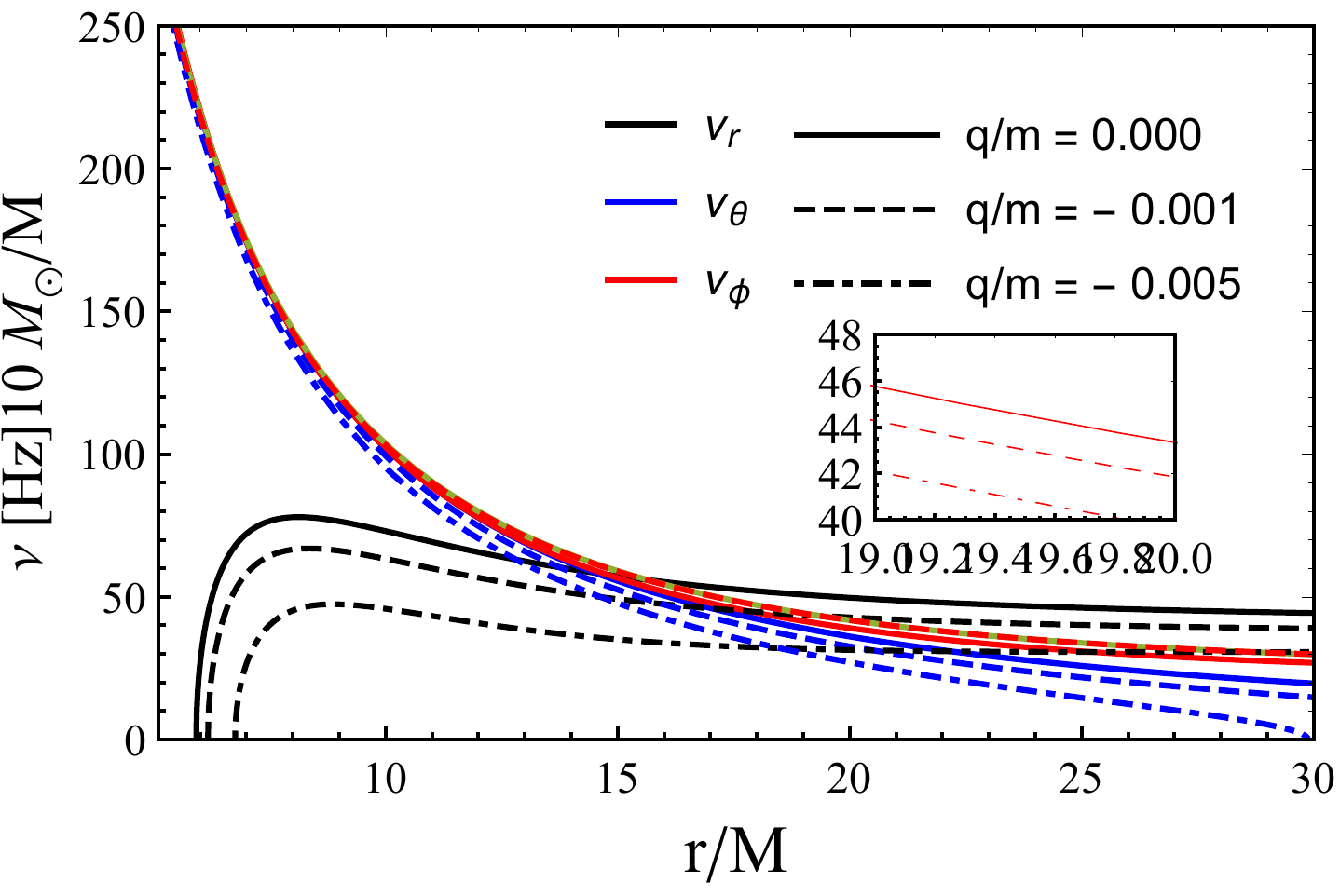}
\end{tabular}
\caption{\label{fig:qpo_freq_charge} Plots show the epicyclic frequencies as the function of $r/M$ in the case with charged particle. Radial, latitudinal and orbital frequencies are plotted for various combinations of charge parameter $\pm q/m$ while keeping fixed magnetic parameter $B=0.005$. }
\end{figure*}

Since the above mentioned frequencies, $\bar{\omega}_{i}$, respectively refer to the locally measured frequencies, one needs to measure these frequencies at infinity for distant observers. Let us then define $\nu$ as frequency measured by distant observer located at infinity. It is worth noticing that the frequencies measured by local and distant observers are related by the following transformation~\cite{Shaymatov20egb,Stuchlik21_qpo}
\begin{eqnarray}
\bar{\omega}\rightarrow\frac{\bar{\omega}}{-g^{tt}\mathcal{E}}\, ,
\end{eqnarray}
which stems from the redshift factor for the transformation 
from the proper time $\tau$ to the time measured at infinity $t$. 
Thus frequencies $\bar{\omega}$ and $\nu_{\phi}$ are related by the following relation with $G$ and $c$
\begin{eqnarray}\label{Eq:rel}
\nu=\frac{1}{2\pi}\frac{c^3}{GM}\frac{\bar{\omega}}{(-g^{tt})\mathcal{E}}\, . 
\end{eqnarray}
Then above frequency can be measured by real observers at infinity can be directly used to analyse the observation data. 

In Fig.~\ref{fig:qpo_freq} we show the epicyclic frequencies as the function of $r/M$ for neutral particle moving around the magnetized Ernst black hole for various combinations of magnetic field parameter $B$. As can be seen from Fig.~\ref{fig:qpo_freq} the radial frequency gets increased and its oscillation shifts left to larger $\nu$ as a consequence of an increase in the value of magnetic field parameter $B$. However for latitudinal frequency it remains almost unchanged while for orbital frequency the magnetic field parameter causes an increase in the value of frequency at larger distances as compared to the one around the Schwarzschild black hole (see Fig.~\ref{fig:qpo_freq}, solid lines). Similarly, in Fig.~\ref{fig:qpo_freq_charge} we present the radial dependence of the epicyclic frequencies of charged particles around the magnetized Ernst black hole for fixed $B$. From Fig.~\ref{fig:qpo_freq_charge},  the impact of parameter $q/m$ gives rise to the increase in the value of all frequencies. {That is the radial frequency significantly increases, while latitudinal and orbital frequencies slightly increase at larger distances when increasing the value of parameter $q/m$. However, the opposite behaviour is the case for the parameter $-q/m$, thus resulting in decreasing values for all epicyclic frequencies as $-q/m$ grows (see Fig.~\ref{fig:qpo_freq_charge}, right panel).} Note that the orbital frequency gets slightly decreased as a consequence of the presence of parameter $-q/m$.  

\section{Constraints on the magnetic field}\label{Sec:Constrain}

In this final section we are interested to put observational constraints on the parameters of the four dimensional magnetized black hole using our theoretical results and the experimental data for three microquasars. In particular,  the appearance of two peaks at 300 Hz and 450 Hz in the X-ray power density spectra of Galactic microquasars has stimulated a lot of  theoretical works to explain the value of the 3/2-ratio \cite{Strohmayer:2001yn}. Although there is no  well-accepted explanation yet, the occurrence of $\nu_U/\nu_L=3/2$ with lower $\nu_L$ Hz QPO, and of the upper $\nu_U$ Hz QPO have been reported for the microquasars  GRO J1655-40, XTE J1550-564 and GRS 1915+105. Bellow we give the corresponding frequencies of the three microquasars \cite{Strohmayer:2001yn}

\begin{multline}\label{pr1}
\text{GRO J1655-40 : }
\nu_U=450\pm 3 \text{ Hz},\;\nu_L=300\pm 5 \text{ Hz},
\end{multline}
\begin{multline}\label{pr2}
\text{XTE J1550-564 : }
\nu_U=276\pm 3 \text{ Hz},\;\nu_L=184\pm 5 \text{ Hz},
\end{multline}
\begin{multline}\label{pr3}
\text{GRS 1915+105 : }
\nu_U=168\pm 3 \text{ Hz},\;\nu_L=113\pm 5 \text{ Hz},
\end{multline}

One possible explanation of the twin values of the QPOs is linked to the so-called phenomenon of resonance. The main idea is that near the vicinity of the ISCO, the in-falling  particles can perform radial as well as vertical oscillations and, in general, the two oscillations couple non-linearly yielding the observed quasiperiodic power spectra \cite{Abramowicz:2003xy,Horak:2006sw}. 
{Note that the frequency ratio $\nu_U/\nu_L$ describes the resonant phenomena for HF QPOs. Thus there exist specific models representing different types of resonances. In the present work, we shall assume that the resonance observed in the three microquasars~\eqref{pr1}, \eqref{pr2} and~\eqref{pr3} is described by the parametric resonance that is given by }
\begin{equation}\label{as1}
\nu_U = \nu_r \mbox{\, \, and\, \,}  \nu_L =\nu_{\theta}.
\end{equation}

To constrain the model we assume a three parameter model for the QPO frequency, and perform a Monte Carlo simulations with a $\chi$-square analysis 
\begin{equation}
\chi^{2}(M,B,r)=\frac{(\nu_{r}-\nu_{1\mathrm{U}})^{2}
}{\sigma^2_{1\mathrm{U}}}\\\notag
+\frac{(\nu_{\theta}-\nu_{1\mathrm{L}})^{2}%
}{\sigma_{1\mathrm{L}}}.
\end{equation}

To simplify the problem further, we have set the charge of the particle to zero. In what follows, we present our results for the constraints on the black hole mass and magnetic field. 
\vspace{1cm}
\begin{itemize}
    \item  Microquasar GRO J1655-40
\end{itemize}
 Within 1$\sigma$ we obtain for the black hole mass $M/\textup{M}_\odot=5.86^{+0.06}_{-0.08}$ and $r/M=6.03^{+0.05}_{-0.02}$. For the magnetic field we obtain $B\sim 2.93 \times 10^{-30} m^{-1}$. Since $B$ in the geometric units has dimension of inverse of length $[L^{-1}]$ %we can find the magnetic field in Guass using $10^{-4}\times G^{1/2} c^{-1} \epsilon_0^{1/2}$. 
 For the particular case, we find for the magnetic field  $B\sim 3.61 \times  10^{-7}$ Gauss. The parametric plot between the magnetic field and the black hole mass is presented in Fig. 8.
 \begin{itemize}
\item Microquasar XTE J1550-564:
\end{itemize}
 Here we find within 1$\sigma$ a black hole mass $M/\textup{M}_\odot=9.42^{+0.19}_{-0.35}$ along with the radii $r/M=6.08^{+0.15}_{-0.06}$. For the magnetic field we obtain $B\sim 4.76 \times 10^{-30} m^{-1}$ which can be written as $B\sim 5.87 \times  10^{-7}$ Gauss. The parametric plot between the magnetic field and the black hole mass is presented in Fig. 9.

\begin{figure}
\includegraphics[width=0.49\textwidth]{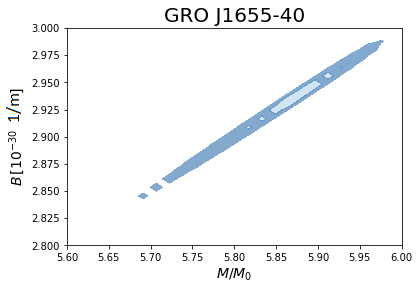}
\caption{The plot shows here the magnetic field $B$ and black hole mass within $68\%$ and $95\%$ CL.}
\end{figure}
\begin{figure}
\includegraphics[width=0.49\textwidth]{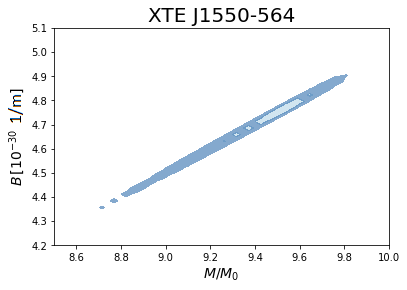}
\caption{The plot shows the parametric plot of magnetic field $B$ and black hole mass within $68\%$ and $95\%$ CL.}
\end{figure}
\begin{figure}
\includegraphics[width=0.49\textwidth]{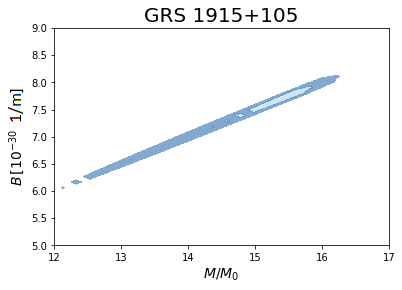}
\caption{The plot shows the parametric plot of magnetic field $B$ and black hole mass within $68\%$ and $95\%$ CL.}
\end{figure}
 
 \begin{itemize}
\item Microquasar GRS 1915+105
\end{itemize}
 In this final case, we find within 1$\sigma$  a black hole mass $M/\textup{M}_\odot=14.10^{+0.68}_{-1.35}$ along with $r/M=6.45^{+0.43}_{-0.20}$. For the magnetic field on the other hand we obtain $B\sim 7.79 \times 10^{-30} m^{-1}$, or alternatively written in Gauss $B\sim 9.61 \times  10^{-7}$ Gauss. The parametric plot for this case can be found in Fig. 10.

\section{Conclusions}
{The recent astrophysical data suggests that accretion disks play a key role in creating the QPOs around super massive black holes and are the primary source to obtain information about gravity and nature of the geometry in the strong field regime and as well as the existing fields surrounding such black holes~\cite{Abramowicz13}.} Also the existing fields play an important role in altering a particle's geodesics, thus strongly influencing observable properties (i.e. the shadow, the ISCO, the QPOs, etc.). With this in view,  the magnetic field is increasingly important in the dynamics of charged particles in the very close vicinity of black holes. Thus, it is worth studying the effect of magnetic fields on the particles moving in the disks around astrophysical black holes.  In this paper we consider interesting solution describing a static and spherically symmetric black hole that includes the additional gravity due to a nonlinear coupling the Schwarzschild black
hole with the Melvin's magnetic universe~\cite{Ernst76}. For these reasons the study of the properties of such solution has value.   

In the present paper we have studied the dynamics of charged particles around the magnetized black hole known as the Ernst black hole. We have found that the radius of the innermost stable circular orbit (ISCO) for both neutral and charged test particles is strongly affected under the effect of magnetic field, thus shrinking its values. The ISCO radius rapidly decreases for the charged test particles. This happens because the magnetic field reflects the combined effects of gravitational and Lorentz forces on the charged particles. Also we have shown that epicyclic frequencies of particles moving around the black hole gets increased significantly as a consequence of the effect of magnetic field.  

In the first part of this work, we have presented the generic form of the epicyclic frequencies and selected three microquasars with known astrophysical quasiperiodic oscilations (QPO) data to constrain the magnetic field. In all three cases, we found that the magnetic field is of the order of magnitude $B\sim  10^{-7}$ Gauss. It is interesting to note that in the present work we have identified the upper frequency with $\nu_r$  and the lower frequency with $\nu_{\theta}$ to explain the QPOs. Usually when the rotation is introduced, people in most of the cases in this resonance model identify the upper frequency with $\nu_{\theta}$ and lower frequency with $\nu_r$, respectively. This suggests that the Ernst metric due to the magnetic field effect which has a backreaction effect on the spacetime can mimic the rotation of the black hole to some extent. This is explained from the fact that when the backreaction effect is considered the spherical symmetry is broken and this leads to very interesting results which can be important from the phenomenological point of view.  Finally, we note that a more realistic estimation of the value of magnetic parameter should include the particle charge. In that case, one must specify the particle, for example it can be an electron, proton or ionize atom and so on having specific $q/m$. Introducing a quantity $c^4/(GM q/m)$, for electrons we get e typical factor of $10^2$, which may suggest an increase of the magnetic field to the magnitude $B\sim 10^{-5}$ Gauss which is consistent with the result found in \cite{Kolos:2017ojf,Stuchlik:2019dlx}.  

{Interestingly, it turns out that the magnetized Reissner-Nordstr\"{o}m black hole solution causes axially symmetric spacetime that is regarded as an analogue of the rotating Ernst spacetime as a consequence of the presence of magnetic field parameter $B$, thereby resulting in mimicking the black hole rotation parameter $a/M$~\cite{Shaymatov21c}. With this in mind, one would therefore be allowed to consider the possible extensions of recent analysis to the case of axially symmetric magnetized black hole spacetime to constrain its parameters, which we would next intend to investigate in a separate work. }

\label{Sec:Conclusion}

\section*{Acknowledgments}
S.S. acknowledges the support from Research F-FA-2021-432 of the Uzbekistan Ministry for Innovative Development. M.J. would like to thank Lim Yen-Kheng for useful comments and discussions related to this work. The work of K.B. was partially supported by the JSPS KAKENHI Grant Number JP21K03547.
\bibliographystyle{apsrev4-1}  %% BibTeX style
\bibliography{gravreferences,ref}

\end{document}